\documentclass[double]{article}
\usepackage{times}
\usepackage{algorithm}
\usepackage{algorithmic}
\usepackage{wrapfig}
\usepackage{verbatim}
\usepackage{amsmath}
\usepackage{graphicx}
\usepackage{subcaption}
\usepackage[active]{srcltx}
\usepackage{color}
\usepackage{lineno}
\usepackage{dirtytalk}
\usepackage{amsthm}
\usepackage{authblk}
\usepackage{listings}
\usepackage{tikz}
\usepackage{xcolor,colortbl}

\definecolor{green}{rgb}{0.1,0.1,0.1}

\usepackage{epsfig,amsfonts}
\usepackage{amsmath,amsthm,amssymb}

\usepackage{adjustbox}
\usepackage{multirow}
\usepackage{setspace}
\usepackage{hyperref}
\usepackage{comment}

\long\def\comment #1\commentend{}

\begin{document}

\title{\Large Can We Mathematically Spot Possible Manipulation of Results in Research Manuscripts Using Benford's Law?}
\author{Teddy Lazebnik$^{1*}$ and Dan Gorlitsky$^{1}$ \\\(^1\) Independent researcher, Israel \\ \(*\) Corresponding author: lazebnik.teddy@gmail.com }

\date{}

\maketitle 

\begin{abstract} 

The reproducibility of academic research has long been a persistent issue, contradicting one of the fundamental principles of science. What is even more concerning is the increasing number of false claims found in academic manuscripts recently, casting doubt on the validity of reported results. In this paper, we utilize an adaptive version of Benford's law, a statistical phenomenon that describes the distribution of leading digits in naturally occurring datasets, to identify potential manipulation of results in research manuscripts, solely using the aggregated data presented in those manuscripts. Our methodology applies the principles of Benford's law to commonly employed analyses in academic manuscripts, thus, reducing the need for the raw data itself. To validate our approach, we employed 100 open-source datasets and successfully predicted $79\%$ of them accurately using our rules. Additionally, we analyzed 100 manuscripts published in the last two years across ten prominent economic journals, with ten manuscripts randomly sampled from each journal. Our analysis predicted a \(3\%\) occurrence of result manipulation with a \(96\%\) confidence level. Our findings uncover disturbing inconsistencies in recent studies and offer a semi-automatic method for their detection.
\end{abstract}
\noindent


\maketitle \thispagestyle{empty}

\pagestyle{myheadings} \markboth{Draft:  \today}{Draft:  \today}
\setcounter{page}{1}

\section{Introduction}
\label{sec:introduction}
The scientific community places great emphasis on maintaining the integrity and dependability of published manuscripts \cite{intro_1,intro_2,intro_3}. The accuracy and validity of research findings are crucial for advancing knowledge and establishing evidence-based policies \cite{intro_4,intro_5}. Unfortunately, the existence of fraudulent or deceptive research across different disciplines presents a substantial obstacle for scientists \cite{intro_6,intro_7,intro_8}.

There are various motivations behind the presentation of misleading results in academic papers. These motivations range from seeking professional recognition by publishing in high-impact journals to securing funding based on impressive previous work, and even attempting to salvage a study that did not yield the desired outcomes \cite{intro_9,intro_10,intro_11}. Furthermore, the traditional peer review process often fails to identify deliberate attempts at result fabrication, particularly when raw data is not provided, although the absence of raw data itself is an undesirable practice \cite{intro_12,intro_13}. This issue is particularly relevant in the field of economics, where data analysis and statistical properties play a crucial role, but restrictions on sharing raw data, driven by privacy concerns and the protection of business secrets, make it difficult to scrutinize the findings \cite{economic_fraud}. Consequently, scholars in this field may find it tempting to manipulate results with minimal risk involved, creating an undesirable environment for research integrity.

Ensuring the integrity and trustworthiness of research studies is essential, and this necessitates the identification and exposure of potential inconsistencies or intentional misrepresentations within research manuscripts \cite{fraud_bad}.  Traditional methods of detecting anomalies or suspicious patterns often involve a manual examination, which is a time-consuming and resource-intensive process \cite{fruad_hard_1,fraud_hard_2}. Furthermore, this approach demands a high level of expertise in each respective field, thereby limiting the number of individuals capable of performing such tasks. As a result, there is an increasing demand for objective and automated approaches to assist in the identification of possible falsehoods in academic research, particularly when the original data is unavailable for review.

This paper presents an innovative method leveraging Benford's law \cite{benford}, a statistical phenomenon commonly utilized in forensic accounting and auditing. Our approach focuses on devising rules for examining standard statistical analyses like mean, standard deviation, and linear regression coefficients. Benford's law centers around the distribution of leading digits in real-world datasets, offering a mathematical framework to detect deviations from anticipated patterns. Building upon this framework, we introduce multiple tests associated with various types of statistical analyses typically reported in research manuscripts. These tests compare the expected Benford's distribution against the observed distribution for each respective analysis.

In order to assess the efficacy of our methodology, a sample of 100 open-access datasets was obtained. For half of these datasets, we computed the actual statistical values, while for the remaining half, we intentionally introduced modifications to these values. The findings demonstrate that our proposed approach successfully predicted the outcomes with an accuracy rate of 79\%. Subsequently, we collected data from 100 papers published in the top 10 economic journals within the last two years. Disturbingly, our method detected anomalies in 3\% of the papers, attaining a confidence level of 96\%.

\begin{figure}[!ht]
    \centering
    \includegraphics[width=0.99\textwidth]{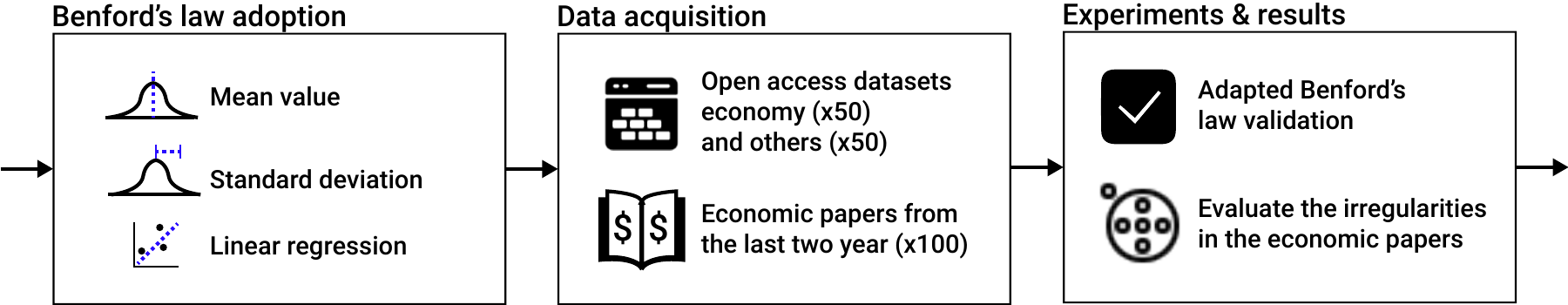}
    \caption{A schematic view of this study. First, we outline the mathematical framework based on Benford's theory. Next, we outline the data acquisition process for the experiments. Finally, we present the experimental setup and results, including a method validation experiment and an evaluation of recent economics studies, followed by an analysis of the results and a discussion about their implementations.}
    \label{fig:flow}
\end{figure}

This paper is organized as follows. Section \ref{sec:model} outlines our adoption of Benford's distribution and the construction of the manuscripts test. Section \ref{sec:experiments} presents the methodology employed to collect and preprocess the data used for our experiments as well as the analysis itself. Section \ref{sec:results}, provides the results of our experiments. Section \ref{sec:discussion} discusses the implementations of our results followed by an analysis of the applications and limitations of the study with possible future work. Fig. \ref{fig:flow} provides a schematic view of this study. 

\section{Statistical Operators Benford's Law}
\label{sec:model}
Benford's law describes the expected distribution of leading digits in naturally occurring datasets \cite{benford}. It states that in many sets of numerical data, the leading digits are not uniformly distributed. In fact, they follow a logarithmic distribution, as follows:
\begin{equation}
    P(d) = log_{10} (1 + \frac{1}{d}),
    \label{eq:benford_basic}
\end{equation}
where \(d \in \{1, 2, \dots, 9\}\) indicates the leading digit and \(P(d) \in [0, 1]\) is the probability a number would have \(d\) as its leading digit. To apply Benford's law in practice, one needs to compare the observed distribution of leading digits in a dataset to these of Eq. (\ref{eq:benford_basic}). Deviations from the expected distribution can indicate potential anomalies, irregularities, or manipulation within the dataset. 

Now, let us consider a set of vectors \(V := \{v_i\}_{i=1}^k \in \mathbb{R}^{n \times k}\). Formally, an irregularity test based on Benford's law would return \(p\) which is the probability value obtained from the Kolmogorov-Smirnov test \cite{kol_siv_test} between the log distribution obtained by fitting \(V\) and the distribution in Eq. (\ref{eq:benford_basic}). In order to perform this test on values obtained from \(V\) using operator \(o\), one needs to first find Benford's distribution associated with such an operator. Hence, let us consider three common statistical operators: mean, standard deviation, and linear regression coefficients. One can numerically obtain these distributions using the convolution operator \cite{method_support}.

Formally, we define an anomaly test to be \(T_{o}(D)\) where \(T_{o}: \mathbb{R}^n \rightarrow [0, 1]\) is a function that accepts a vector \(D \in \mathbb{R}^n\) and an operator \(o\) and returns a score of the probability \(D\) is anomaly with respect to operator \(o\). Formally, for our case, we associate each operator \(o\) with its Benford's distribution and \(T_{o}(D)\) is implemented to return \(1 - p\) where \(p\) is the probability value obtained from the Kolmogorov-Smirnov test \cite{kol_siv_test} between the distribution associated with the operator \(o\) and the same one after fitting to \(V\). Notably, for each operator, we generated 1000 random samples and calculated the results for each one of them. We denoted the worst result obtained as \(a \in [0, 1]\). In order to ensure that the proposed test numerically produces results in the range \([0, 1]\), for each outcome, \(x\), we compute and report \((x-a)/(1-a)\).

\section{Experimental Setup}
\label{sec:experiments}
In this section, we outline the two experiments conducted in this study. The first experiment is designed to numerically validate the performance of the proposed method. After validating the method, in a complementary manner, the second experiment evaluates the number of irregularities in recent academic economic studies. We implemented the experiments using the Python programming language \cite{python} (Version 3.7.5). We set \(p < 0.05\) to be statistically significant. 

First, for the method's performance validation, we manually collect 100 numerical datasets from the Data World\footnote{We refer the reader to  \url{https://data.world/datasets/economics}} and Kaggle\footnote{\url{https://www.kaggle.com}}, following \cite{teddy_datasets}. The datasets are randomly chosen from a broad range of fields and represent a wide range of computational tasks. Each dataset is represented by a matrix \(D\). We define a feature \(f_j\) of a dataset \(D\) as follow \(f_j := \forall i \in [1, \dots, n]: d_{i,j}\). A feature is used to calculate the unitary statistical properties. Based on this data, for each datasets \((D)\) and statistical operator \((o)\), we computed \(T_o(D)\), obtaining a vector of results denoted by \(u\). The overall anomaly probability prediction is define to be \(\frac{1}{|u|}\sum_{i = 1}^{|u|}u_i\). For half of the datasets, we introduce uniformly distributed noise which is between 1 and 10 percent of the mean value in a uniform manner. As such, these datasets should not agree with Benford's law and therefore if the proposed method predicts they do, it is an error. As such, we have 50 positive and 50 negative examples. 

Second, for the manuscript evaluation, we collected a sample of 100 papers published in 10 leading economic journals over the past two years. These papers served as the test subjects for applying our proposed method to detect anomalies or irregularities. We choose these amounts and distribution to balance the time and resource burden and the statistical power of the sample. In order to determine which journals are leading in the economics field, we used the Scimago Journal and Country rank website\footnote{We refer the reader to \url{https://www.scimagojr.com/}}, searching for the \say{Economics and Econometrics} and taking the top 10 journals \cite{scimago_1,scimago_2,scimago_3}: Quarterly Journal of Economics, American Economic Review, Journal of Political Economy, Journal of Finance, Review of Economic Studies, Econometrica, Journal of Economic Literature, Review of Financial Studies, Journal of Marketing, and Journal of Financial Economics. For each journal, we mainly count how many manuscripts the journal published in the last two years, asking the computer to randomly pick 10 indexes. Once the indexes were obtained, we downloaded these manuscripts from the journals' websites. Next, we manually extract the results from the manuscripts presented either in tables or figures. For each of them, if appropriate, we apply our adopted Benford's law. 

\section{Results}
\label{sec:results}
To assess the performance of our method, we evaluated the confusion matrix for the dataset, as presented in Table \ref{table:confusion}. The obtained results indicate an accuracy of 0.79 and an \(F_1\) score of 0.77. Notably, the model exhibited a tendency to predict manipulation-free manuscripts incorrectly, identifying 7 manipulation-free manuscripts as containing manipulations. Conversely, it also misclassified 14 manuscripts with manipulations as manipulation-free. However, from the perspective of the journal, it is preferable for the model to err on the side of caution by falsely predicting manuscripts as manipulation-free, as falsely accusing innocent authors of result manipulation is deemed more undesirable than missing manuscripts with actual manipulations.

\begin{table}[!ht]
\centering
\begin{tabular}{l|cc|c}
                  & \textbf{Positive}          & \textbf{Negative}          & \textbf{Total} \\ \hline
\textbf{Positive} & \cellcolor[HTML]{9AFF99}43 & \cellcolor[HTML]{FFCCC9}7 & 50             \\
\textbf{Negative} & \cellcolor[HTML]{FFCCC9}14 & \cellcolor[HTML]{9AFF99}36 & 50             \\ \hline
\textbf{Total}    & 57                      & 43                      & 100           
\end{tabular}
\caption{A confusion matrix of the proposed method on the open source datasets.}
\label{table:confusion}
\end{table}

Furthermore, Table \ref{table:manuscripts} provides an overview of the predicted number of economic manuscripts flagged for containing results manipulations based on varying confidence levels. It is evident that as the confidence level increases, the number of flagged manuscripts decreases. This observation aligns with expectations since the null hypothesis assumes that the manuscripts are manipulation-free. Hence, higher confidence levels necessitate stronger statistical evidence of manipulation for a manuscript to be flagged.

\begin{table}[!ht]
\centering
\begin{tabular}{l|ccccc}
\textbf{Confidence level}    & 90\% & 92\% & 94\% & 96\% & 98\% \\ \hline
\textbf{Flagged manuscripts} & 12   & 8    & 6    & 3    & 2   
\end{tabular}
\caption{The number of manuscripts flagged to have irregularities according to our method as a function of the required confidence level. }
\label{table:manuscripts}
\end{table}

\section{Discussion and Conclusion}
\label{sec:discussion}
In this study, we introduced an innovative approach to identify potential falsehoods in research manuscripts by applying Benford's law to commonly reported statistical values, including mean, standard deviation, and linear regression coefficients. By adopting this law to the context of research manuscripts, we aimed to enhance the detection of deceptive information.

To validate the efficacy of our approach, we conducted two experiments. In the initial experiment, we evaluated the performance of our method by applying it to a random sample of 100 datasets from diverse fields. The results demonstrated that our method achieved an accuracy of 0.79 and an F1 score of 0.77, indicating its capability to identify potential anomalies, albeit with some limitations. Consequently, it can serve as a supportive tool or an initial filter to alleviate the burden of manual investigation.

Building upon this premise, the second experiment involved applying our method to 100 recent manuscripts from reputable high-impact academic journals in the field of economics. Alarming findings emerged, as approximately 3\% of the manuscripts exhibited anomalies, inaccuracies, or even explicit manipulations, with a confidence level of 96\%. These outcomes unfortunately align with existing trends in academic fraud practices, underscoring the significance of our approach in uncovering inconsistencies and deliberate misrepresentations in academic research. 

By leveraging Benford's law, our method offers an objective and automated solution to complement traditional manual scrutiny. Furthermore, it holds particular relevance in fields like economics where researchers heavily rely on data analysis and statistical properties, often lacking access to raw data due to privacy or proprietary constraints. While our results demonstrate the promise of our approach, there are several limitations to consider. First, our method relies on the assumption that the reported aggregated data follows Benford's distribution, which may not always hold true \cite{limit_1_1,limit_1_2}. Second, our approach requires the development of tests for each statistical operator, which makes it hard to utilize a wide spectrum of fields and manuscripts that may use and report about a large number of different statistical analysis methods. Third, our method does not provide definitive proof of fraud or misconduct but rather serves as a signal for potential irregularities that warrant further investigation, thus only slightly reducing the time and resources required for the task. Finally, the publication of this study reduces its effectiveness as malicious scholars would be aware of the proposed method and develop counter-strategies to overcome it, as common in other fields like cybersecurity \cite{final}.




 
\bibliography{biblio}

\begin{thebibliography}{10}

\bibitem{intro_1}
M.~Franzen.
\newblock {\em Science Between Trust and Control: Non-Reproducibility in
  Scholarly Publishing}, chapter~22, pages 467--485.
\newblock John Wiley \& Sons, Ltd, 2016.

\bibitem{intro_2}
D.~Fanelli.
\newblock Do pressures to publish increase scientists' bias? an empirical
  support from us states data.
\newblock {\em PLOS ONE}, 5(4):1--7, 2010.

\bibitem{intro_3}
H.~Lee, S.~E. Lamb, M.~K. Bagg, E.~Toomey, A.~G. Cashin, and G.~L. Moseley.
\newblock Reproducible and replicable pain research: a critical review.
\newblock {\em PAIN}, 159(9):1683--1689, 2018.

\bibitem{intro_4}
C.~C. Lewis, S.~Fischer, B.~J. Weiner, C.~Stanick, M.~Kim, and R.~G. Martinez.
\newblock Outcomes for implementation science: an enhanced systematic review of
  instruments using evidence-based rating criteria.
\newblock {\em Implementation Science}, 10:155, 2015.

\bibitem{intro_5}
P.~Roberts and H.~Priest.
\newblock Reliability and validity in research.
\newblock {\em Nursing Standard}, 20(44), 2006.

\bibitem{intro_6}
R.~M. Frederickson and R.~W. Herzog.
\newblock Keeping them honest: Fighting fraud in academic publishing.
\newblock {\em Molecular Therapy}, 29(3):889--890, 2021.

\bibitem{intro_7}
J.~A.~T. da~Silba and Q-H. Vuong.
\newblock Do legitimate publishers benefit or profit from error, misconduct or
  fraud?
\newblock {\em Exchanges: The interdisciplinary Research Journal}, 8(3), 2021.

\bibitem{intro_8}
J.~Gu, X.~Wang, C.~Li, J.~Zhao, W.~Fu, G.~Liang, and J.~Qiu.
\newblock Ai-enabled image fraud in scientific publications.
\newblock {\em Patterns}, 3(7):100511, 2022.

\bibitem{intro_9}
D.~D. Bergh, B.~M. Sharp, H.~Aguinis, and M.~Li.
\newblock Is there a credibility crisis in strategic management research?
  evidence on the reproducibility of study findings.
\newblock {\em Strategic Organization}, 15(3):423–436, 2017.

\bibitem{intro_10}
L.~M. Chambers, C.~M. Michener, and T.~Falcone.
\newblock Plagiarism and data falsification are the most common reasons for
  retracted publications in obstetrics and gynaecology.
\newblock {\em BJOG: An International Journal of Obstetrics \& Gynaecology},
  126(9):1134--1140, 2019.

\bibitem{intro_11}
J.~Brainard and J.~You.
\newblock What a massive database of retracted papers reveals about science
  publishing's ‘death penalty', 2018.
\newblock Accessed on May 29th, 2023.

\bibitem{intro_12}
D.~P. Misra and V.~Ravindran.
\newblock Peer review in academic publishing: threats and challenges.
\newblock {\em J R Coll Physicians Edinb}, 49:99--100, 2019.

\bibitem{intro_13}
J.~Kelly, T.~Sadeghieh, and K.~Adeli.
\newblock Peer review in scientific publications: Benefits, critiques, \& a
  survival guide.
\newblock {\em EJIFCC}, 25(3):227--243, 2013.

\bibitem{economic_fraud}
S.~F. Karabag and C.~Berggren.
\newblock Retraction, dishonesty and plagiarism: Analysis of a crucial issue
  for academic publishing, and the inadequate responses from leading journals
  in economics and management disciplines.
\newblock {\em Journal of Applied Economics and Business Research},
  2(3):172--183, 2012.

\bibitem{fraud_bad}
W.~E. Stehbens.
\newblock Basic philosophy and concepts underlying scientific peer review.
\newblock {\em Medical Hypotheses}, 52(1):31--36, 1999.

\bibitem{fruad_hard_1}
M.~Cokol, F.~Ozbay, and R.~Rodriguez-Esteban.
\newblock Retraction rates are on the rise.
\newblock {\em EMBO reports}, 9(1), 2008.

\bibitem{fraud_hard_2}
F.~C. Fang and A.~Casadevall.
\newblock Retracted science and the retraction index.
\newblock {\em Infection and Immunity}, 79(10):3855--3859, 2011.

\bibitem{benford}
C.~Durtschi, W.~Hillison, and C.~Pacini.
\newblock The effective use of benford's law to assist in detecting fraud in
  accounting data.
\newblock {\em Journal of Forensic Accounting}, pages 17--34, 2004.

\bibitem{kol_siv_test}
F.~J. Massey.
\newblock The kolmogorov-smirnov test for goodness of fit.
\newblock {\em Journal of the American Statistical Association},
  46(253):68--78, 1951.

\bibitem{method_support}
D.A. Nix and A.S. Weigend.
\newblock Estimating the mean and variance of the target probability
  distribution.
\newblock In {\em Proceedings of 1994 IEEE International Conference on Neural
  Networks (ICNN'94)}, volume~1, pages 55--60, 1994.

\bibitem{python}
K.~R. Srinath.
\newblock Python – the fastest growing programming language.
\newblock {\em International Research Journal of Engineering and Technology},
  4(12), 2017.

\bibitem{teddy_datasets}
T.~Lazebnik, T.~Fleischer, and A.~Yaniv-Rosenfeld.
\newblock Benchmarking biologically-inspired automatic machine learning for
  economic tasks.
\newblock {\em Sustainability}, 2023.

\bibitem{scimago_1}
M.~E. Falagas, V.~D. Kouranos, R.~Arencibia-Jorge, and D.~E. Karageorgopoulos.
\newblock Comparison of scimago journal rank indicator with journal impact
  factor.
\newblock {\em The FASEB Journal}, 22(8):2623--2628, 2008.

\bibitem{scimago_2}
A.J. Gomez-Nunez, B.~Vargas-Quesada, Z.~Chinchilla-Rodríguez, V.~Batagelj, and
  F.~Moya-Anegon.
\newblock Visualization and analysis of scimago journal \& country rank
  structure via journal clustering.
\newblock {\em Aslib Journal of Information Management}, 68(5):607--627, 2016.

\bibitem{scimago_3}
J.~Manana-Rodríguez.
\newblock {A critical review of SCImago Journal \& Country Rank}.
\newblock {\em Research Evaluation}, 24(4):343--354, 2014.

\bibitem{limit_1_1}
A.~Shukla, A.~K. Pandey, and A.~Pathak.
\newblock Benford’s distribution in extrasolar world: Do the exoplanets
  follow benford’s distribution?
\newblock {\em Journal of Astrophysics and Astronomy}, 38:7, 2017.

\bibitem{limit_1_2}
L.~Barabesi, A.~Cerioli, and D.~Perrotta.
\newblock Forum on benford’s law and statistical methods for the detection of
  frauds.
\newblock {\em Statistical Methods \& Applications}, 30:767--778, 2021.

\bibitem{final}
J.~S. Granick.
\newblock {The Price of Restricting Vulnerability Publications}.
\newblock {\em Intl. J. CommLaw \& Pol'y}, 9, 2005.

\end{thebibliography}
\bibliographystyle{unsrt}

\end{document}